\documentstyle[preprint,aps] {revtex}
\def\slash#1{\setbox0=\hbox{$#1$}#1\hskip-\wd0\hbox to\wd0{\hss\sl/\/\hss}}
\begin{document}
\draft
\title{Dynamical Chiral Symmetry Breaking, Goldstone's Theorem and the
Consistency of the Schwinger--Dyson and Bethe--Salpeter Equations
}

\author{H.J. Munczek}
\address{Department of Physics and Astronomy, University of Kansas,
Lawrence, Kansas 66045}
\date{October, 1994}
\maketitle
\begin{abstract}
A proof of Goldstone's theorem is given for the case in which global chiral
symmetry is dynamically broken. The proof highlights a needed consistency
between the exact Schwinger--Dyson equation for the fermion propagator and the
exact Bethe--Salpeter equation for fermion--antifermion bound states. A
criterion, based on the Cornwall, Jackiw and Tomboulis effective action for
composite operators, is provided for maintaining the consistency when the
equations are modified by approximations. For gauge theories in which partial
conservation of the axial current (PCAC) should hold, a constraint on the
approximations to the fermion--gauge boson vertex function is discussed, and a
vertex model is given which satisfies both the PCAC constraint and the vector
Ward--Takahashi
identity.
\end{abstract}

\section{Introduction and Main Result}
A large amount of work on dynamical chiral symmetry breaking and its
application to low energy strong interaction physics [1-4] suggests a picture
of
the low mass pseudoscalar mesons as ``almost" Nambu--Goldstone (NG)
bosons. The NG bosons would arise due to the spontaneous breaking of the flavor
chiral
symmetry
present in the usual quark model when the weak and electromagnetic interactions
are
absent. For light quarks the small masses induced by the Higgs mechanism have
been included in the formalism quite successfully by partially conserved
current algebra, operator product, and renormalization group techniques which
further
support the above picture[2].

An alternative to those methods is to calculate the properties
of the mesons as quark--antiquark bound states through the use of the
Bethe--Salpeter
(BS)
equation, which can be derived rigorously from the underlying field theory and
therefore preserves its symmetries. In this context another equation that
plays an essential role is the Schwinger-Dyson (SD) equation for the fermion
propagator, a quantity which appears explicitly in the BS equation.
 Spontaneous chiral symmetry breaking is signaled by the
appearance of an otherwise absent scalar term in the quark propagator.
Consequently
this change in the propagator should
be reflected in the BS equation through the appearance of a massless
pseudoscalar
solution. Since the early work of Nambu and Jona--Lasinio[1] this has been
found
to happen explicitly in a variety of chiral invariant models when use is made
of the ladder approximation {\it both} for the SD
and BS equations[1,5]. Also in the ladder approximation it has been found
numerically[6] that when
a small quark mass term is added to the the SD equation, the successful results
of current algebra and other general techniques mentioned above are preserved.

It is obvious that the ladder approximation to the SD equation for a fermion is
not entirely satisfactory. The equation has been the subject of extensive
research, particularly of studies directed to incorporate an improved structure
for the vertex
function, in order for it to
satisfy the vector Ward--Takahashi (WT) identity and to exhibit correct
infrared and ultraviolet behaviors. These studies include the use of the gauge
technique[7] as well as the use of algebraic combinations of fermion
propagators[8]. A natural
consistency requirement which then arises is the following: given a change in
the
SD equation which
takes
it beyond the ladder approximation, the BS equation should also
change in a way that
preserves the appearance of NG bosons and other dynamically broken chiral
symmetry features.

A procedure to guarantee this consistency requirement is given in what follows
by using the effective action
formalism for composite operators developed by Cornwall, Jackiw and Tomboulis
(CJT), who showed that it can be conveniently applied to the analysis of the SD
equation and of dynamical symmetry breakdown[9]. It has also been shown that
the
formalism is a natural framework for the exact treatment of the S--matrix for
bound
states, the Bethe--Salpeter equation and related Green functions[10]. The CJT
formalism will be applied here to give a proof of Goldstone's theorem for the
specific
case of dynamical chiral symmetry breaking in a theory with fermions. The proof
is constructive in that it emphasizes the common origin of the SD and
BS equations and thus provides a prescription to maintain the validity of the
theorem and related chiral symmetry features when subjecting the equations to
truncations and other approximations.

I start by
considering a situation in which there is a set of fermions $\psi_a(x)$ in
interaction with some other fields, such that one can construct a CJT action
$\Gamma[B]$ which is a functional of a ``classical" bilocal, bispinor, field
$B_a^b(xy)$, with each of the labels $a,b$ indicating spinor as well as
internal symmetry indices. In what follows, the indices will
occasionally be left implicit. Repeated variables and indices are assumed to be
integrated or summed over.
The CJT action yields[9] the exact SD equation for the fermion
propagator $S_F(x,y)$
\begin{equation}
{\delta\Gamma \over \delta B(xy)}\bigg|_{B=S_F}=0\; \; ,
\end{equation} where
$$iS_F(x,y)=<0\mid T\psi(x)\bar\psi (y)\mid 0>\; \; .\eqno (1a)$$
One can also see[10,11] that the exact BS equation for a bound state of mass
$M$ described
by a wave function
\begin{equation}
\psi_a^b(x,y,p)=\chi_a^b (x-y,p)e^{ip(x\alpha+y(1-\alpha))}\; ,
\end{equation}
where $0<\alpha<1$, and $p^2=M^2$, has the form
\begin{equation}
{\delta^2\Gamma [B]\over \delta B_a^b(xy)\delta
B_{a'}^{b'}(x'y')}\bigg| _{B=S_F} \psi_{a'}^{b'}(x',y',p)=0\; .
\end{equation}
The action is
now assumed to be obtained from a formally global chiral invariant lagrangian
field theory in which the fermions transform as
\begin{equation}
\psi'(x)=e^{i\gamma_5\theta\tau^\ell}\psi(x),\; \; \;  \bar\psi'(y)=\bar\psi
(y)
e^{i\gamma_5\theta\tau^\ell}\;.
\end{equation}
The parameter $\theta$
is real and the $\tau^\ell$ are hermitian matrix representations of the
generators of
the flavor group. I also assume that there are no anomalies in the axial
currents
associated with the indices $\ell$ in Eq. (4). Under those circumstances
inspection of the CJT action shows that
\begin{equation}
\Gamma[B']=\Gamma[B]\; ;\; \; \; \; B'(x,y)=e^{i\gamma_5\theta\tau^\ell}
B(x,y)e^{i\gamma_5\theta\tau^\ell}\; \; .
\end{equation}
Under an infinitesimal
transformation with $\theta= \epsilon$ we have then
\begin{equation}
0=\delta_5\Gamma[B]= {\epsilon\delta\Gamma\over\delta B_{a'}^{b'}(x',y')}
\{i\gamma_5\tau^\ell, B(x',y')\}_{a'}^{b'} \; .
\end{equation}
 From Eq. (6) we
also obtain
\begin{equation}
0={\delta\over\delta B_a^b(xy)} \bigg[\delta_5\Gamma[B]\bigg]\; ,
\end{equation}
or in more detail,
\begin{equation}
0={\delta^2\Gamma[B]\over \delta B_a^b(x,y)\delta B_{a'}^{b'}(x'y')}
\{\gamma_5\tau^\ell, B(x'y')\}_{a'}^{b'} + {\delta\Gamma\over \delta
B_{a'}^b (xy)} (\gamma_5\tau^\ell)_{a'}^a+ {\delta\Gamma\over \delta
B_a^{b'}(xy)} (\gamma_5\tau^\ell)^{b'}_b \; .
\end{equation}
 Setting $B(xy)=S_F
(x,y)$
in (8), the last two terms vanish if the SD equation (1) holds. We see then
that
if the vacuum is not chiral invariant, that is if
\begin{equation}
\{\gamma_5\tau^\ell, S_F(x,y)\}=<0\mid\{\gamma_5\tau^\ell, T \psi
(x)\bar\psi(y)\}\mid
0>\not= 0\; ,
\end{equation}
 then the BS equation has a pseudoscalar solution
of vanishing four--momentum, a Nambu--Goldstone boson, since equation (8)
becomes
\begin{equation}
{\delta^2\Gamma[B]\over \delta B_a^b(xy)\delta
B_{a'}^{b'}(x'y')}\bigg| _{B=S_F}\{ \gamma_5 \tau^\ell, S_F
(x',y')\}_{a'}^{b'}=0 .
\end{equation}
The results described above are
true for the exact SD and BS equations. It is clear that any approximate
treatment of either equation has to be accompanied by a treatment of the other
equation which maintains the validity of Goldstone's theorem[12]. From the
derivation above we see that this will happen if both the approximated SD and
BS
equations are derived through equations (1) and (3) from the same approximated,
but
chiral invariant,
bilocal effective action satisfying equation (5).
\newpage

\section{Formalism and Applications}
%\noindent A. {\it Basic Definitions and Chiral Properties}
\subsection{\it Basic Definitions and Chiral Properties}

I now consider the specific case of the fermions interacting with
vector gauge fields $A_\mu$. The action for such a system can be
written as
\begin{equation}
S(\psi,\bar\psi,A_\mu)=\int \bigg\{\bar\psi (i \slash\partial_x-m)\psi +{\cal
L}(A_\mu)+{\cal L}_I(\psi,\bar\psi,A_\mu)\bigg\}d^4x\; .
\end{equation}
 ${\cal
L}(A_\mu)$ is the lagrangian for the gauge fields and includes gauge fixing
terms and ghost fields, when present. The $A_\mu$ fields do not change under
global or local chiral transformations. The fermions transform as given by Eq.
(4), and the interaction lagrangian ${\cal L}_I (\psi,\bar\psi,A_\mu)$ is
assumed to be invariant under those transformations.

The CJT action $\Gamma[B]$ can be obtained as follows. $Z[J]$, the generating
functional for fermion Green functions, is given by
\begin{equation}
Z[J]={1\over N}\int D\psi D\bar\psi DA_\mu Exp\bigg\lbrack i
S(\psi,\bar\psi,A_\mu)-i\int \bar\psi(x)J(xy)\psi(y)dxdy\bigg\rbrack
\end{equation}
where $Z[0]=1$. Then, $W[J]$, the generating functional for connected fermion
Green's functions, can be expressed by
\begin{equation}
W\lbrack J\rbrack =-i\ell n Z\lbrack J\rbrack\; .
\end{equation}
The classical
bilocal field $B(x,y)$ is defined as
\begin{equation}
B(x,y)\equiv -i\delta W\lbrack J\rbrack/\delta J(yx)\; \; ,
\end{equation}
with
$$B(xy)\bigg|_{J=0}=-i<0\mid T\psi(x)\bar\psi(y)\mid0>=S_F(xy)\; .\eqno
(14a)$$
Finally, the effective action is constructed as
\begin{equation}
\Gamma\lbrack B\rbrack =W\lbrack J\rbrack - i B(xy)J(yx)\; \; .
\end{equation}
{}From (15) we see that
$$i\delta\Gamma[B]/\delta B(xy)=J(yx)\; ,\eqno (15a)$$ and, because of(14a),
the SD
equation (1) follows.

For the system of fields analyzed here, the effective action has the form[9]
\begin{equation}
\Gamma\lbrack B\rbrack =-i Tr\lbrace (i\slash\partial-m)B\rbrace
+\bar\Gamma[B]\; \;,
\end{equation}
 where $\bar\Gamma\lbrack
B\rbrack$ is invariant under local as well as global chiral transformations.
If we perform an infinitesimal transformation
\begin{equation}
\delta_5
B(xy)=i\gamma_5\tau^\ell\epsilon(x)B(xy)+B(xy)i\gamma_5\tau^\ell\epsilon(y)\;
\; ,
\end{equation}
Eq. (16) yields the basic expression
\begin{equation}
Tr\bigg\lbrace{\delta\Gamma\over\delta B}\delta_5B\bigg\rbrace=-i Tr\lbrace
(i\slash\partial -m)\delta_5
B\rbrace\; \; .
\end{equation}

This equation can also be obtained directly from expression (12) by performing
an\break infinitesimal chiral change in the integration variables
and by use of the
definitions (13)--(15).

If the chiral transformation is global and if the mass matrix $m$ is zero, the
right hand side in Eq. (18) vanishes. Then Goldstone's theorem follows in the
manner discussed in section I.

If the chiral transformation is local, and because the infinitesimal $\epsilon
(x)$
is otherwise an arbitrary function of $x$, an integration by parts in (18)
yields

\begin{eqnarray}
tr\int & \bigg\lbrack &\tau^\ell\gamma_5\gamma_\cdot \partial_x\lbrace
B(xy)\delta(x-y)\rbrace +i\lbrace m,\tau^\ell\rbrace\gamma_5
B(xy)\delta(x-y)\bigg\rbrack dy\\ \nonumber
& - &tr\int\bigg\lbrace{\delta\Gamma\over\delta B(xy)}\tau^\ell\gamma_5
B(xy)+{\delta\Gamma\over \delta B(yx)}
B(yx)\gamma_5\tau^\ell\bigg\rbrace dy=0\; .\\ \nonumber
\end{eqnarray}

In this expression $tr$ indicates trace over discrete indices and there is no
integration over $x$.
Eq. (19) expresses, in the framework of the effective action, the contents of
the partially conserved axial current (PCAC)
relationships among Green functions. For instance, taking the functional
derivative of (19) with respect to $B_a^b(x'y')$, setting $B=S_F$, and applying
the
resulting operator to a pseudoscalar solution of the BS equation results in
the exact relationship
\begin{equation}
tr\bigg\lbrack \tau^\ell\slash p\gamma_5\chi(0,p)\bigg\rbrack=tr\bigg\lbrack
\lbrace m,\tau^\ell\rbrace\gamma_5\chi(0,p)\bigg\rbrack\; ,
\end{equation}
where use has been made of Eqs. (1), (2) and (3).

 Eq. (20) can also be obtained through the use of field operator methods
and gives [2], in first order in $m$, the PCAC formula of Gell--Mann, Oakes and
Renner
for pseudoscalar masses[13]. Since the ladder approximation SD and BS equations
can
be obtained from an approximated CJT effective action of the form of Eq. (16)
with $\bar\Gamma [B]$ locally chiral invariant, these equations should give
solutions satisfying PCAC conditions like that of Eq. (20). As discussed in
Section I, this is the case both qualitatively and quantitatively.

\bigskip
%\noindent B. {\it SD Equation and Fermion--Antifermion--Gauge Boson Vertex}
\subsection{\it SD Equation and Fermion--Antifermion--Gauge Boson Vertex}

In Eq. (16) $\bar\Gamma[B]$ can be written as the sum of free and interacting
parts, each one locally chiral invariant, in the form[9]
\begin{equation}
\bar\Gamma[B]=i\; Tr\; \ell n\; B+\Gamma_2[B]\; \; .
\end{equation}
 We then have, with
implicit discrete indices,
\begin{equation}
\delta\Gamma/\delta B(yx)=-iS^{-1}_0(x,y)+iB^{-1}(xy)+\delta
\Gamma_2/\delta B(yx)\; ,
\end{equation}
$$S_0^{-1}(x,y)=-i\slash\partial_y \delta(x-y)-m\delta(x-y)\; ,\eqno (22a)$$
 where $S_0^{-1}(x,y)$ is the free inverse
propagator. The SD equation is, setting $B=S_F$,
\begin{equation}
-S_F^{-1}(x,y)+S_0^{-1}(x,y)+i\; \delta \Gamma_2/\delta
B(yx)\Bigg|_{B=S_F}=0\;.
\end{equation}
For simplicity, one can restrict the discussion to the case in
which there is only one gauge vector field. The ``self--mass" term in Eq. (23)
can be written in the symmetrized form
\begin{eqnarray}
\Sigma(x,y) & = &i{\delta \Gamma_2\over \delta B(yx)}\bigg|_{B=S_F}=-i{1\over
2}g^2\int d^4zd^4x'\bigg\lbrace
\gamma_\mu G_{\mu\nu}(x,z)S_F(x,x')\Gamma_\nu(z;x',y)\\ \nonumber
& + &\Gamma_\mu(z;x,x')S_F(x',y)G_{\mu\nu}(z,y)\gamma_\nu\bigg\rbrace,\\
\nonumber
\end{eqnarray}
where both the vector boson propagator $G_{\mu\nu}$ and the vertex function
$\Gamma_\nu$ are functionals of $S_F$.
 Since
$\Gamma_2[B]$ is invariant under a local chiral transformation, Eq. (24) shows
that the self--mass $\Sigma(x,y)$ and the
vertex function $\Gamma_\nu(z;x,y)$ should transform like $S_F^{-1}(x,y)$, that
is[14]
\begin{equation}
\Sigma(x,y)\rightarrow e^{-i\gamma_5\tau^\ell\theta(x)}\Sigma(x,y)
e^{-i\gamma_5\tau^\ell\theta(y)}\;,
\end{equation}
$$
\Gamma_\nu(z;x,y)\rightarrow e^{-i\gamma_5\tau^\ell\theta(x)}
\Gamma_\nu(z;x,y)e^{-i\gamma_5\tau^\ell\theta(y)}\; ,
\eqno (25a)$$
under the substitution
\begin{equation}
S_F(x,y)\rightarrow e^{i\gamma_5\tau^\ell\theta(x)}
S_F(x,y)e^{i\gamma_5\tau^\ell\theta(x)} .
\end{equation}
 This requirement is
compatible with the vector Ward--Takahashi (WT) identity
\begin{equation}
{\partial\over\partial z_\mu}\Gamma_\mu (z;x,y)=i\bigg\lbrace
\delta(y-z)-\delta(x-z)\bigg\rbrace S_F^{-1}(x,y)\; .
\end{equation}

The vertex function $\Gamma_\mu(z;x,y)$ satisfies its own SD equation which
couples it to higher order Green's functions, thus making it necessary to use
approximation methods. A large number of models[3,7,8] have been discussed for
$\Gamma_\mu(z;x,y)$ with the requirement that it satisfy the WT identity (27)
as well as appropriate symmetry properties and renormalization requirements.
Many of the models, usually presented in momentum space, are linear in
$S_F^{-1}(p)$ and $S^{-1}_F(p+k)$, which appear multiplied by functions of $p$
and $p+k$, the momenta of the fermions at the vertex. In configuration space
this involves derivatives with respect to $x$ and $y$ and therefore, those
models generally
fail to satisfy the condition (25) and the PCAC equation (19), since those
models imply an additional explicit local chiral symmetry breaking in the
effective action.

A vertex model linear in $S_F^{-1}$ and satisfying Eqs. (23) and (25a) can be
constructed as
\begin{equation}
\bigg\lbrack \Gamma_\mu(z;x,y)\bigg\rbrack_a^b=\bigg\lbrack F_\mu(y-z;
x-z)\bigg\rbrack_{aa'}^{bb'}\bigg\lbrack S^{-1}_F(x,y)\bigg\rbrack_{b'}^{a'}\;
\;,
\end{equation} where the matrix structure of $F_\mu$ is such that it preserves
condition (25a). $F_\mu$ should also satisfy
\begin{equation}
{\partial\over\partial z_\mu}\bigg\lbrack F_\mu(z-y;
z-x)\bigg\rbrack^{bb'}_{aa'}=i\bigg\lbrack\delta (y-z)-\delta(x-z)\bigg\rbrack
\delta_a^{b'}\delta_{a'}^b\; \; .
\end{equation}
As an example, a simple choice for $F_\mu$ is
\begin{equation}
F_\mu=\delta_a^{b'}\delta_{a'}^b\int \bigg\lbrack e^{iq\cdot (z-y)}-e^{iq\cdot
(z-x)}\bigg\rbrack
{(x_\mu-y_\mu)\over q\cdot (x-y)}\; \; {d^4q\over (2\pi)^4}+F_\mu^T\; ,
\end{equation} where $\partial F^T_\mu/\partial z_\mu =0$.
In momentum space, with
\begin{equation}
\Gamma_\mu(z;x,y)={1\over (2\pi)^8}\int e^{-ip'\cdot
(x-z)-ip\cdot(z-y)}\Gamma_\mu (p',p)d^4p'd^4p,
\end{equation}
and
\begin{equation}
S^{-1}_F(x-y)={1\over (2\pi)^4}\int e^{-ip\cdot(x-y)}S_F^{-1}(p)d^4p,
\end{equation} the form (28) with the choice (30) translates into[15]
\begin{equation}
\Gamma_\mu(p+k,p)={\partial\over\partial p_\mu}\int^1_0 S^{-1}_F (p+\alpha
k)d\alpha+ \Gamma_\mu^T\; ,
\end{equation} which, with $k_\mu\Gamma_\mu^T=0$,  satisfies the WT identity
\begin{equation}
k_\mu\Gamma_\mu(p+k,p)=S^{-1}_F(p+k)-S^{-1}_F(p).
\end{equation}
\bigskip

%\noindent C. {\it SB Equation}
\subsection{\it BS Equation}

As shown in Section I, taking a further derivative with respect to $B$ in Eq.
(22) and
then setting $B=S_F$ allows one to write the BS equation (3). Approximations to
the exact action $\Gamma[B]$ which maintain the chiral symmetry properties of
the system can be obtained from the loop expansion[9] of the CJT action or,
alternatively, by approximating the self--mass and vertex functionals in such a
way that (25) and (25a) hold. If we make the assumption that their functional
dependence on B is the same as their dependence on $S_F$ we can define for
use in the BS equation
\begin{equation}
{\delta^2\Gamma_2[B]\over \delta B(x_1y_1)\delta
B(xy)}\bigg|_{B=S_F}\equiv{\delta\Sigma(x,y)\over \delta S_F(x_1,y_1)}\; .
\end{equation}
If an explicit form is not readily available for $\Gamma_2[B]$ the chiral
properties can be probed ``on--shell", that is for $B=S_F$,
by using the SD equation (23), which can be written as
$\delta\Gamma/\delta S_F(x,y)=0$. We then have, for $m=0$, the identity
\begin{equation}
0=\epsilon\bigg\lbrace i\gamma_5\tau^\ell, {\delta\Gamma[S_F]\over\delta
S_F(x,y)}\bigg\rbrace\; .
\end{equation}
Using (22), (25), (26) and (35) we see that (36) is equivalent to
\begin{equation}
0=\delta_5\bigg\lbrace{\delta\Gamma[S_F]\over \delta S_F(x,y)}\bigg\rbrace
=i\epsilon {\delta^2\Gamma[S_F]\over \delta S_F(x,y)\delta
S_F(x_1,y_1)}\{\gamma_5\tau^\ell, S_F(x_1y_1)\}\; ,
\end{equation} which demonstrates Goldstone's theorem. If instead of (36) we
consider the identity
\begin{equation}
0=i\gamma_5\tau^\ell\epsilon (y){\delta\Gamma[S_F]\over\delta
S_F(x,y)}+{\delta\Gamma[S_F]\over\delta S_F(x,y)}
i\gamma_5\tau^\ell\epsilon(x)\; ,
\end{equation} we obtain
\begin{equation}
i\gamma_5\tau^\ell[\epsilon(y)-\epsilon(x)]
\slash\partial_x\delta(x-y)+\delta(x-y)\epsilon(x)
\{\gamma_5\tau^\ell,m\}+{\delta^2\Gamma[S_F]\over\delta  S_F(x,y)\delta
S_F(x_1,y_1)}
\delta_5 S_F(x_1,y_1)=0\; .
\end{equation}
Applying this
expression to a pseudoscalar solution $\chi(x-y,p)e^{ip(x\alpha+y(1-\alpha))}$
of the BS equation the
last term vanishes, and we have again, as in Eq. (20), the PCAC relationship
\begin{equation}
tr[\tau^\ell\slash p\gamma_5\chi(0,p)-\gamma_5\{\tau^\ell,m\}\chi(0,p)]\int
d^4x\epsilon(x)e^{ipx}=0\; \; .
\end{equation}

\section{Conclusions}

The results discussed in Section II.A. show that the CJT formalism is a very
convenient tool for the study of bound state fermion--antifermion equations and
their chiral symmetries, both global and local. The formalism provides a very
close connection between the exact SD and BS equations which allows a
straightforward
proof of Goldstone's theorem and the study of the effects of the explicit
breaking of chiral symmetry (PCAC). In addition, the formalism provides a
procedure to maintain the chiral consistency of approximated SD and BS
equation.

In II.B it was seen that the PCAC constraint Eq. (25) on the local chiral
properties of the vertex function, namely that it transforms as the inverse
fermion propagator, can be implementated, along with the WT identity,  when
modeling the vertex function. The
constraint is present in quantum electrodynamics and chromodynamics when
studying spontaneous chiral symmetry breaking and its dependence on the
coupling
constant. The use, in this context, of vertex models that do not satisfy the
PCAC constraint is therefore open to question.

Finally, it was shown in II.C. that if a satisfactory model is given for the
vertex function, and therefore for the self--mass function, then ``on--shell"
approximated functionals  $\delta\Gamma/\delta S$ and $\delta^2\Gamma/\delta
S\delta S$ can be
defined. The functionals give SD and BS equations which are chirally
compatible,
yielding Goldstone's theorem and the consequences of PCAC discussed in II.A.

\begin{acknowledgements}
The author thanks P. Jain, D.W. McKay, S. Pokorski and J.P. Ralston for useful
comments. The
work was supported in part by the U.S. Department of Energy under grant number
DE--FG02--85--ER40214.
\newpage

%\begin{references}
\centerline{\bf REFERENCES}

\noindent [1] Y. Nambu and G. Jona--Lasinio, {\it Phys. Rev.} {\bf 122}, 345
(1961); M.
Baker, K. Johnson, and B.W. Lee, {\it ibid}, {\bf 133}, B209 (1964); H. Pagels,
{\it Phys. Rev. D} {\bf 7}, 3689 (1973); R. Jackiw and K. Johnson, {\it ibid},
{\bf 8}, 2386 (1973); J.M. Cornwall and R.E. Norton, {\it ibid}. {\bf 8}, 3338
(1973); E.J. Eichten and F.L. Feinberg, {\it ibid.} {\bf 10}, 3254 (1974). More
recent references and some well known definitions, equations and identities
mentioned in the present paper can be found in [2], [3] and [4] below. %The
%metric used here is that of [2] and [4].

\noindent [2] V.A. Miransky, {\it Dynamical Symmetry Breaking in Quantum Field
Theories}
(World Scientific, Singapore, 1993).

\noindent [3] C.D. Roberts and A.G. Williams, in {\it Progress in Particle and
Nuclear
Physics}, {\bf 33}, 477 (1994) (Pergamon, Great Britain, 1994).

\noindent [4] S. Pokorski, Gauge Field Theories (Cambridge, University Press,
Great Britain, 1987).

\noindent [5] R. Delbourgo and M.D. Scadron, {\it J. Phys. G} {\bf 5}, 1621
(1979).

\noindent [6] H.J. Munczek and P. Jain,
{\it Phys. Rev. D} {\bf 46}, 438 (1992).

\noindent [7] R. Delbourgo and A. Salam {\it Phys. Rev.} {\bf 125}, 1398
(1964);
R. Delbourgo  and P. West, {\it J. Phys.} {\bf A10}, 1049 (1977); {\it Phys.
Lett.} {\bf 72B}, 96 (1977); R. Delbourgo {\it Nuovo Cim.} {\bf 49A}, 484
(1979). J.E. King, {\it Phys. Rev.} {\bf D27}, 1821 (1983). P. Rembiesa, {\it
Phys. Rev.} {\bf D33}, 2333 (1986). B. Haeri, Jr. {\it Phys. Rev.} {\bf D43},
2701 (1991).

\noindent [8] J.S. Ball and T.W. Chiu, {\it Phys. Rev.} {\bf D22}, 2542 (1980).
D.C. Curtis and M.R. Pennington, {\it Phys. Rev.} {\bf D42}, 4165 (1990). D.
Atkinson, J.C.R Bloch, V.P. Gusynin, M.R. Pennington and M. Reenders, {\it
Phys. Rev. Lett.} {\bf B329}, 117 (1994). Z. Dong, H.J. Munczek and C.D.
Roberts {\it Phys. Lett. B}, {\bf 333}, 536 (1994). A. Bashir and M.R.
Pennington, preprint, Durham Center for Particle Theory, DTP-94/98.

\noindent [9] J.M. Cornwall, R. Jackiw and E. Tomboulis, {\it Phys. Rev.} {\bf
D10}, 2428 (1974).

\noindent [10] D.W. McKay and H.J. Munczek, {\it Phys. Rev.} {\bf D40}, 4151
(1989).

\noindent [11] R. Fukuda, {\it Prog. Theor. Phys.} {\bf 87}, 1487 (1987).

\noindent [12] J. Goldstone, {\it Nuovo Cimento} {\bf 19}, 154 (1961). J.
Goldstone, A. Salam and S. Weinberg, {\it Phys. Rev.} {\bf 127}, 962 (1962).

\noindent [13] M. Gell--Mann, R.J. Oakes and B. Renner, {\it Phys. Rev.} {\bf
175}, 2195 (1968).

\noindent [14] Since $S^{-1}_F(x,x')S_F(x',y)$=$\delta(x-y)$, we have that
$\delta_5 S^{-1}_F(x,y)$=$[\delta S^{-1}_F(x,y)/\delta S_F(x',y')]\times$
$\delta_5 S_F(x',y')$=--$S^{-1}_F (x,x')\delta_5 S_F(x',y')S_F^{-1}(y',y)=
-\{i\gamma_5\tau^\ell\epsilon (x)
S_F^{-1}(x,y)$+$S_F^{-1}(x,y)i\gamma_5\tau^\ell \epsilon(y)\}$, which shows
that
$S^{-1}_F(x,y)\rightarrow
e^{-i\gamma_5\tau^\ell\theta(x)}S^{-1}_F(x,y)e^{-i\gamma_5\tau^\ell\theta(y)}$.

\noindent [15] H.J. Munczek, {\it Phys. Lett.} {\bf B175}, 215 (1986). In this
reference an expression identical to Eq. (33), but with $S^{-1}_F(p+\alpha k)$
replaced by $-S_F(p+\alpha k)$, was given for the dressed vertex
$\Lambda_\mu(p+k,p)=S_F(p+k)\Gamma_\mu(p+k,p)S_F(p).$

\end{document}